\documentclass[aps,prb,twocolumn,
showpacs]{revtex4}
\usepackage{ae}
\usepackage[T1]{fontenc}
\usepackage[ansinew]{inputenc}
\usepackage{amsmath}
\usepackage{amssymb}
\usepackage{graphicx}
\usepackage{color}
\usepackage[colorlinks]{hyperref}
\usepackage{epstopdf}

\begin{document}

\date{\today}



\title{Bound states induced giant oscillations of the conductance in the quantum Hall regime}

\author{A. M. Kadigrobov$^{1}$ and M.V. Fistul$^{1,2}$}
\affiliation{$^{1}$ Theoretische Physik III, Ruhr-Universit\"at
Bochum, D-44801 Bochum, Germany \\
$^{2}$ National University of Science and Technology MISIS, Moscow 119049, Russia}

\begin{abstract}
We theoretically studied the quasiparticle transport in a 2D
electron gas biased in the quantum Hall regime and in the presence
of a lateral potential barrier. The lateral junction hosts the
specific magnetic field dependent quasiparticle states highly
localized in the transverse direction. The quantum tunnelling
across the barrier provides a complex bands structure  of a
one-dimensional
 energy spectrum of these bound states, $\epsilon_n(p_y)$, where $p_y$ is the electron momentum in
 the longitudinal direction $y$. Such a spectrum manifests itself by a large number of peaks and
 drops in the  dependence of the magnetic edge states transmission coefficient  $D(E)$ on the electron
 energy $E$.  E.g., the high value of $D$ occurs as soon as the electron energy $E$  reaches
 gaps in the spectrum. These peaks and drops of $D(E)$ result  in  giant oscillations of the
  transverse conductance $G_x$ with the magnetic field and/or the
 transport
   voltage. Our theoretical
  analysis based on the coherent macroscopic quantum superposition of the bound states
and the magnetic edge states propagating along the system
boundaries, is in a good accord with the experimental observations found in Ref. \cite{kang}

\end{abstract}


 \maketitle

 Great interest has been devoted to  theoretical and experimental studies of  various low-dimensional
 systems such as tunnel junctions, quantum point contacts (QPC), quantum nanowires, and  2D
  electron gas based nanostructures,  just to name a few. These systems show a large variety
  of fascinating quantum-mechanical effects  on the macroscopic scale, e.g. the weak localization
   \cite{Heinzel,Dittrich}, the quantum Hall effect \cite{QH}, the macroscopic quantum tunnelling
   \cite{Devoret}, the conductance quantization in QPCs \cite{Heinzel,QH,Beenakker} etc.

The quantum-mechanical dynamics of quasiparticles and the
electronic transport become even
  more complex and intriguing when
 \emph{bound states} are present in such systems.
  Indeed, it was shown in Ref. \cite{ARinQPC1,ARinQPC2} that the bound states manifest
  themselves by narrow drops (the anti-resonance) in the gate voltage dependent conductance
  of QPCs. It is well known also that the bound states naturally arising on the boundaries
  of various systems such as graphene nanoribbons \cite{Castro},  nanowires \cite{Charlier,Mourik},
  2D electron gas under magnetic field, greatly influence the transport properties.

Specific magnetic field dependent bound states can be artificially
created if a lateral junction (barrier) is fabricated inside of a
2D electron gas subject to an externally applied magnetic field
(see Fig. 1) \cite{exp1,kang,twin,barrier,grapheneLat}.  These
highly localized
 states are formed due to the quantum-mechanical  interference of magnetic edge states occurring
 on the both sides of the barrier.   The localization length
 in the transverse direction $x$ is
of the order of the magnetic
length $\ell_c~\simeq~\sqrt{\hbar c/(eH)}$ in the regime of a strong externally applied
magnetic field, $H$.

However, in the longitudinal direction $y$ these states show
delocalized behavior, and the dynamics of electrons is
characterized by the component of electron momentum $p_y$. Thus,
in such a 1D channel the  energy  spectrum of electrons,
$\epsilon_n(p_y)$, contains many bands, and it is shown for two
lowest bands in Fig. 2A. In the absence of quantum tunnelling
across the junction the left-right symmetry of the electron states
results in a large number of degenerate states in the electronic
spectrum (see Fig. 2A, dashed line). The quantum tunnelling
provides the lifting of the degeneracy, and the energy spectrum
shows a complex structure  with bands and gaps (see Fig. 2A, solid
line). Notice here that in the region of a rather small magnetic
field such a spectrum was calculated in Ref. \cite{twin,barrier},
and in the region of a strong magnetic field, i.e in the quantum
Hall regime, it was obtained in Ref. \cite{kang}. Since this
spectrum having an origin in the coherent quantum-interference
phenomenon,  is extremely sensitive to various interacting
effects, the renormalization of the above-mentioned spectrum  due
to the Coulomb interaction and /or the impurities has been
theoretically studied in papers
\cite{Mitra,Kollar,Nonoyama,Kim,Aranzana}.

  \begin{figure}
 \centerline{\includegraphics[width=0.7\columnwidth]{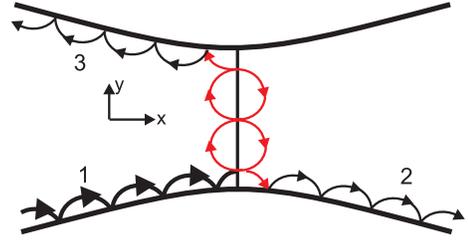}}
 \caption{\textbf{The  point contact with a lateral junction in the quantum Hall regime.}
 A scattering process (propagation and reflection) of magnetic edge states on the electron
  states localized on the lateral junction, is schematically shown.}
 \label{edgestates}
 \end{figure}

The bands and gaps in the energy spectrum of electronic states
bounded to the junction, directly manifest themselves  in  a large
amount of interesting effects as the transport in the longitudinal
direction is studied. Indeed, the giant oscillations of the
longitudinal conductance with the gate voltage, strongly nonlinear
current-voltage
 characteristics and coherent Bloch oscillations under a weak
 electric fields, have been
 predicted and theoretically studied \cite{twin,barrier}.

It is  naturally to suppose that the above-mentioned  unique
electronic spectrum can be probed by the electric current flowing
\emph{across } a lateral junction. In this article we show that the
bands and gaps in the quasiparticle spectrum of the states bounded
to a lateral junction,
 manifest themselves in  giant oscillations of the conductance
 $G_x$ as the applied magnetic
  field or dc voltage are varied. The origin of these oscillations is
  in the magnetic edge
  states transmission coefficient $D(E)$ which shows  peaks and drops as
  the electron energy is tuned.

Qualitatively, the quantum-mechanical dynamics of electrons in a
strong magnetic field and in the presence of a lateral junction
can be considered as following: the lateral junction serves as a
tunable quantum-mechanical scatterer for  propagating edge states
(see Fig. 1). E.g., the edge state $1$ propagating from the left
lead along the lower boundary is scattered by the barrier into the
edge state $2$ going to the right lead,  and into the edge state
$3$ which reflects from the barrier to the left lead along the
upper boundary. The highest probability of such reflection
 occurs if the propagation of electrons \emph{along} the junction is allowed. It takes place if
 the electron energy $ E$ is inside of the energy bands of bound states. In this case the
 transmission coefficient $D(E)$ shows minimal values. As the energy of electrons is tuned to
 gaps in the spectrum of the bound states, the edge states propagation along the junction  is
 forbidden, and a great enhancement of $D(E)$ is obtained. These oscillations of $D(E)$ transform
 in  giant oscillations of $G_x$ under variation of
 the  magnetic field or the dc voltage. Such oscillations of
 $G_x$  have been verified experimentally in  the Ref. \cite{kang} but
 the quantitative analysis has  not been  done. Notice here,
 that in the Ref.\cite{Takagaki}, the electron current flowing perpendicular to
the lateral junction in a 2D electron gas under a strong magnetic
filed was numerically calculated, and similar oscillations of the
conductance were obtained.

Below we present a complete analytical solution of the transverse
transport of electrons in such a system that allows to clarify
both qualitatively and quantitatively experimental features
\cite{kang} such as  the small value of the observed conductance
in comparison to the expected Landauer conductance, the
oscillations of the conductance as a function of the magnetic
field or voltage, the dependence of the conductance on the lateral
size of the system, etc. Our analysis based on the
quantum-mechanical scattering of propagating magnetic edge states
on the bound states is in a good accord with both the qualitative
scenario and experimental observations.

Let us consider a QPC fabricated in a two-dimensional electron gas
in the presence of a lateral junction and subject to an externally
applied magnetic field. The magnetic field $H$ is perpendicular to the QPC plane.
The QPC is characterized by the coordinate-dependent electrostatic potential
$\widetilde V (x, y)=\frac{m \omega_1^2}{2}y^2 + V (x)$,
%
%
%
%
where  the last term describes the potential barrier between two
parts of the electron gas. The parameter $\omega_1$ characterizes
the curvature of the confinement potential of the QPC, $m$ is the
electron effective mass.

Quantum dynamics of electrons in the QPC with the lateral junction is
described by the wave function $\Psi(x,y)$ satisfying
the two-dimensional Schr\"odinger equation:

%
%
\begin{eqnarray}
 &-&\frac{\hbar^2}{2 m} \frac{\partial^2 \Psi}{\partial x^2}+
 \Big[\frac{1}{2 m}\big(-i \hbar \frac{\partial }{\partial y}-
 \frac{e H x}{c}\big)^2 \nonumber \\
 &+&\frac{m \omega_1^2}{2}y^2 + V (x)-E\Big]\Psi =0,
\label{Shcroedinger}
\end{eqnarray}
%
%
where the Landau  gauge, i.e. the vector-potential  ${\bf A} =
(0,Hx,0)$, is used. Here, the axis $y$ is parallel to the barrier
and the $x$-axis is directed along the QPC (see Fig. 1).
Introducing the dimensionless variables, $\xi=x/\ell_c$ and
$\varepsilon = E/(\hbar \omega_c)$ (the $\omega_c$ is the
cyclotron frequency), we write the total wave function $\Psi$ as
%
%
\begin{eqnarray}
 \Psi(\xi,y)= \int_{-\infty}^\infty Q(p_y,\xi)\exp\{i\frac{ p_y y }{\hbar}\}d p_y, \nonumber \\
 Q(p_y,\xi)=\sum_{n=0}^{\infty} R_n(p_y)\varphi_{n,p_y}(\xi)
\label{Fourier}
\end{eqnarray}
Here, the partial wave functions $\varphi_{n,p_y}(\xi)$ describe  the magnetic field dependent electron states bounded to the lateral junction and satisfy to the equation
\begin{eqnarray}
  \frac{\partial^2 \varphi_{n,p_y}}{\partial \xi^2}-
  \Big[\Big(\frac{p_y \ell_c}{\hbar} -\xi \Big)^2+
 \nu (\xi)-2 \varepsilon_{n}(p_y) \Big]\varphi_{n,p_y}(\xi) =0
\label{Shcroedinger2}
\end{eqnarray}
%
%
with the boundary conditions  being $\varphi_{n,p_y}(\xi)
\rightarrow 0$ at $\xi \rightarrow \pm \infty$. Here, the
dimensionless potential of the junction is $\nu (\xi) = 2m V(\xi
l_c) \ell_c^2/\hbar^2$. These bound states are formed from the
edge states propagating along the barrier
 $\phi_{n,p_y}^{(l)}(\xi)$ ($\phi_{n,p_y}^{(r)}(\xi)$) on the left (right) parts of
the lateral junction. In the quantum Hall regime as $E~\simeq~\hbar \omega_c$ these bound
 states decay in the transverse direction  on the distance $\ell_c$. In the absence of
 quantum tunneling the electronic spectrum contains a large amount of degenerate states. The quantum-mechanical interference between  left and right edge states results in a lifting of this degeneracy, and
a peculiar one-dimensional spectrum $\varepsilon_n(p_y)$ with an
alternating  sequence of narrow energy bands $\sim
\sqrt{1-|t|^2}\; \hbar \omega_c$ and energy gaps,
\cite{twin,kang,barrier} $\Delta_n \sim |t|\;\hbar \omega_c$,
where $|t|^2$ is the barrier transparency (see Fig. 2A).

The transverse electronic transport is determined by the  partial wave
functions $R_n (p_y)$ which, in turn, are entangled with the bound
states. Next, we explicitly consider this quantum entanglement in
the case that the energy of electrons is in the vicinity of the
first "crossing point". This crossing point occurs at  $p_y = 0$,
and the energy $\epsilon_0=3/2$ (in dimensionless units).

The generic total quantum-mechanical state is presented as a
superposition of basis functions, i.e.
\begin{eqnarray}
Q_0(p_y,\xi)=C_1(\varepsilon)R^{(l)}_0 (p_y)\phi^{(l)}_{0, p_y}(\xi)
 +C_2(\varepsilon)R^{(r)}_0(p_y)\phi^{(r)}_{0, p_y}(\xi)
\label{Superposition}
\end{eqnarray}
Here, $R^{(l,r)}_0 = R_0 \pm R_1$ are the functions normalized to
the unit flux density, the energy dependent coefficients
$C_1(\varepsilon)$ and $C_2(\varepsilon)$ determine the
transmission coefficient $D(E)$ and, therefore, the electronic
transport.

The partial wave functions $R_0^{(l,r)}(p_y)$ are satisfied to the
following set of coupled  equations (its derivation is given in
Supplementary Materials):
%
%
\begin{eqnarray}
\left(\frac{\alpha^2}{2} \frac{\hbar^2}{\ell_c^2} \frac{d^2 }{d
p_{y}^2} + \varepsilon_0^{(l)}(p_y) -\varepsilon \right)R_0^{(l)}
+
\tilde{\Delta}R_0^{(r)}  =0;\nonumber \\
\left(\frac{\alpha^2}{2} \frac{\hbar^2}{\ell_c^2} \frac{d^2 }{d
p_y^2} + \varepsilon_0^{(r)}(p_y) -\varepsilon \right)
  R_0^{(r)}  +\tilde{\Delta} R_0^{(l)} =0,
\label{BasicEquation}
\end{eqnarray}
%
%
where $\alpha =\omega_1 /\omega_c $, and
$\varepsilon_0^{(l,r)}(p_y)$ are the energy spectrum of the left(
right) edge states in the absence of tunneling (dashed line in
Fig. 2A), the dimensionless energy gap
$\tilde{\Delta}=\Delta/(\hbar \omega_c)$ is determined by the
quantum tunneling across the barrier.

As  $\alpha \ll 1$  one may solve  this set of equations in the quasiclassical approximation.
 Indeed, substituting  $\Psi_{1,2} =A_{1,2}(p_y)\exp\{i S(p_y)\ell_c/(\hbar\alpha)\}$ in
 Eq.(\ref{BasicEquation}) and introducing the classical momentum as $P =d S/dp_y$ one
 finds that the electron dynamics is determined by the quasiclassical phase trajectories
 $P(p_y)$ as
%
%
\begin{eqnarray}
P^2=\varepsilon - \frac{\varepsilon_0^{(l)}(p_y)+\varepsilon_0^{(r)}(p_y)}{2} \pm \nonumber \\
\pm \sqrt{\Big(\frac{
\varepsilon_0^{(l)}(p_y)-\varepsilon_0^{(r)}(p_y)}{2}\Big)^2+\tilde{\Delta}^2}.
\label{momentum}
\end{eqnarray}
%
%
Notice here that the quasiclassical approximation is valid as
$|\varepsilon-\varepsilon_0^{(l,r)}(p_y)| \gg \alpha^{2/3}$.

The quasiclassical phase trajectories determined by
Eq.(\ref{momentum}) for different values of the electron energy
$\varepsilon$ are shown in Fig. 2B. As the well separated phase
trajectories approach to each other the quantum tunneling  occurs
between them. However, in  contrast to the standard interband
transitions \cite{LandauZienerBlount}, in the case under
consideration the tunnelling takes place in the vicinity of
Lifshitz's phase transition \cite{Lifshitz}. Indeed, as it follows
from Eq.(\ref{momentum}) there are two critical energies at which
the topology of trajectories changes: at first, in the vicinity of
the energy $\varepsilon_{cr}^{(1)} = \varepsilon_0
-\tilde{\Delta}$ the mutual directions of the motion on two open
trajectories vary (see Fig. 2Ba and 2Bb); secondly, a new closed
orbit arises at  the critical energy $\varepsilon_{cr}^{(2)} =
\epsilon_0 +\tilde{\Delta}$ (see Fig. 2Bc and 2Bd).
The quasiclassical approximation breaks down not only at the turning points but also
in the vicinity of the point $p_y=0$ where the quantum tunneling takes place.

  \begin{figure}
 \centerline{\includegraphics[width=0.7\columnwidth]{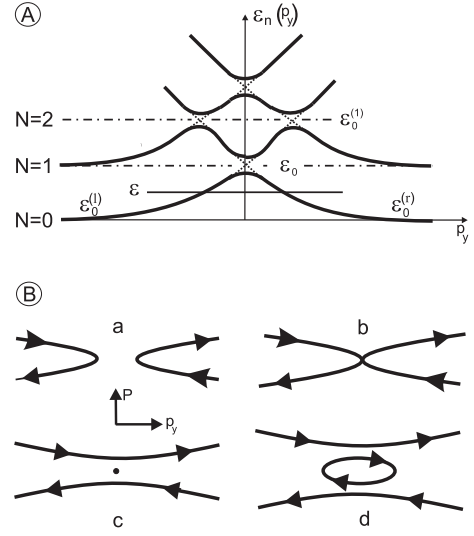}}
 \caption{\textbf{The energy spectrum of electron states bounded to the lateral
 junction and corresponding quasiclassical trajectories.}
A) The lowest two bands  of the  spectrum $\epsilon_n(p_y)$  are
shown by dashed
  (in the absence of tunneling) and solid (in the
  presence of  tunneling) lines.
 B) The phase trajectories $P(p_y)$  for different values of the electron energy:
 a)$\varepsilon<\varepsilon_{cr}^{(1)}=\varepsilon_0-\tilde{\Delta}$;
 b)$\varepsilon=\varepsilon_{cr}^{(1)}$;
  c) $\varepsilon=\varepsilon_{cr}^{(2)}=\varepsilon_0+\tilde{\Delta}$;
   d)$\varepsilon>\varepsilon_{cr}^{(2)}$ (here $2 \tilde{\Delta}$ is the energy gap).}
 \label{trajectories}
 \end{figure}

In order to properly elaborate quantum tunnelling between various
quasiclassical phase trajectories we apply  the perturbation
analysis which is valid in a whole ranges of energies but the
energy gap $\tilde{\Delta}$ is assumed to be small. Introducing
the Fourier transformation as
%
%
\begin{eqnarray}\label{Fourier2}
\Phi(p_y)_{1,2}=\int_{-\infty}^{\infty} g_{1,2}(Y)
 \exp \{i\frac{Y p_y \ell_c }{\hbar\alpha}\} dY
\end{eqnarray}
%
%
and presenting  the Fourier transform $g_{1,2}(Y)$ in the form
%
%
\begin{eqnarray}\label{u}
g_{1,2}=u_{1,2}(\zeta)e^{\{\pm i(\frac{2}{3}\zeta^3+2 \eta \zeta)\}},
\end{eqnarray}
%
%
one gets the following set of equations \cite{PRL} for the new
variable $\zeta=Y[\frac{\ell_c \omega_c}{\alpha v}]^{1/3}$ (here,
$v=|d\varepsilon_0^{(l)}(p_y)/dp_y|$ is the electron velocity):
%
\begin{eqnarray}\label{SystemH3}
&&\hspace{5mm} i\frac{du_1(\zeta)}{d\zeta}=-
\gamma e^{-i\left(\frac{2}{3}\zeta^3 +
2\eta \zeta\right)} u_2(\zeta) \nonumber \\
&&\hspace{5mm} i\frac{du_2(\zeta)}{d\zeta}=+
\gamma e^{i\left(\frac{2}{3}\zeta^3 + 2\eta\zeta\right)} u_1(\zeta),
\end{eqnarray}
where the parameters
%
%
%
%
%
%
\begin{eqnarray}
\gamma = 2^{1/3} \tilde{\Delta} \Big[\frac{\ell_c \omega_c}{\alpha
v}\Big]^{2/3}, \hspace{3mm} \eta = 2^{1/3}(\varepsilon_0
-\varepsilon)\Big[\frac{\ell_c \omega_c}{\alpha v} \Big]^{2/3}.
\label{Parameters}
\end{eqnarray}
%
%
Here, the parameter $\eta$ controls  the topology of the phase
trajectories (see Fig.\ref{trajectories}B) and the parameter
$\gamma$ determines the probability of quantum tunnelling between
trajectories.

The perturbation analysis of Eq.(\ref{SystemH3}) in the limit of
$\gamma \ll 1$ allows one to obtain the probability $D(E)$ for a
particle in the momentum space  to be transmitted from $p_y
\rightarrow + \infty $  to $p_y \rightarrow - \infty $ as
\begin{eqnarray}\label{probability2}
D(\epsilon) =  \gamma^2 \Big[\pi 2^{2/3} \textrm{Ai} \left( 2^{2/3}
\eta \right) \Big]^2,
\end{eqnarray}
where $\textrm{Ai}(x)$ is the Airy function \cite{Stegan} (see
Supplementary Material, Section II). The typical dependence of
$D(\epsilon)$ is shown in Fig. 3.

  \begin{figure}
 \centerline{\includegraphics[width=0.7\columnwidth]{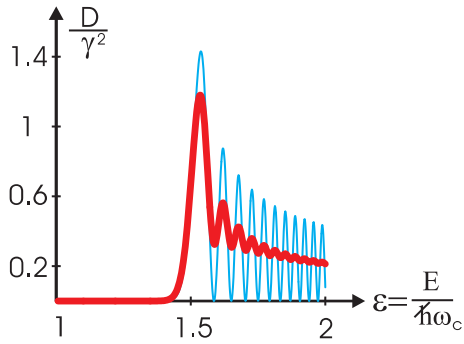}}
 \caption{\textbf{Energy dependent transmission probability.}
 Dependence of the transmission probability $D$ on the normalized
 energy $\varepsilon$ is   shown. The transparency of  the lateral barrier (which acts as a
 tunable quantum scatterer) is zero as long as the electron energy is below the lowest energy
 gap at $E_0/\hbar \omega_c = 1.5$ (see Fig. \ref{trajectories}A). Further increase
 of $E$ results in a narrow sharp peak accompanied by fast
 transparency oscillations (thin solid line) caused by the quantum
 interference inside the effective scatter.  The fast oscillations are  partially washed out at finite
 temperature $T$ (thick solid line).   The parameters $\alpha=10^{-2}$, $v/(l_c \omega_c)= 2$,
 and $k_B T/\hbar \omega_c=10^{-2}$  have been used. }
 \label{edgestates}
 \end{figure}
One can see that  $D(E)$ is an extremely small as $E<(3/2)\hbar
\omega_c$. In the opposite
 regime
at $E>(3/2)\hbar \omega_c$,  $D(E)$ displays both a great
enhancement and fast oscillations
 with
a small period of $\delta E_1~\simeq~\hbar \omega_c \alpha^{2/3}$.
The origin of these oscillations is the quantization of a
quantum-mechanical phase of the wave function of electrons moving
along the closed orbit  (see, Fig. 2Bd). At finite temperatures
these fast oscillations  are partially washed out, and it results
in a strong decay of the function $D(E)$ as  $E>(3/2)\hbar
\omega_c$ (see Fig. 3, thick solid line).

As we turn to the electronic transport in the quantum Hall regime
it can be shown (see Supplementary Material,  for details) that
$D(E)$ determines also the energy  dependent transmission
coefficient of electrons propagating across the lateral junction.
Thus,  $D(\mu)$ determines the linear conductance $G_x$. Here,
$\mu$ is the chemical potential of the system.

The current-voltage characteristic $I(V)$ of the QPC in a broad voltage region is obtained by
making
use of the Landauer-B\"{u}ttiker approach \cite{QH} as:

\begin{eqnarray}\label{current}
I(V) = \frac{2 e}{h} \sum_{n}\int d E D_n(E) \\\nonumber
[f(E+eV/2)- f(E-e V/2)]~,
\end{eqnarray}
where $f(E)$ is the Fermi-Dirac distribution function. Notice
here, that the transmission coefficients for upper bands, $D_n$,
are also determined by the energy differences to the crossing
points $\varepsilon_n$ as
$$
D_n=\gamma^2_n \Big[\pi 2^{2/3} \textrm{Ai} \left( 2^{2/3} \eta_n
\right) \Big]^2
$$
where $$\eta_n \approx 2^{1/3}(\varepsilon_n
-\varepsilon)[\frac{\ell_c \omega_c}{\alpha v_n}]^{2/3}$$ and
$$\gamma_n \approx 2^{1/3}\Delta_n \Big[\frac{\ell_c
\omega_c}{\alpha v_n}\Big]^{2/3}$$ (see Supplementary Materials,
Section III, for details) . The differential conductance $G_x$ is
determined as $G_x=dI/dV$. The typical dependence of $G_x(V)$
determined by Eqs. (\ref{probability2}) and (\ref{current}) is
shown in Fig. 4.

The dependence of $G_x(V)$ displays giant oscillations with the period
$\delta E_2~\simeq~\hbar \omega_c$. The two reasons determine the appearance of
peaks in the $G_x(V)$ dependence: at first, the applied voltage distorts the spectrum of bound
states to $\epsilon_n(p_y, eV)$, and  secondly, the voltage tunable energy of electrons $E=\mu+eV/2$
traces the energy gaps $\Delta_n$ in the spectrum of bound states.

The consistent experimental study of the conductance of a 2D
electron gas biased in the quantum Hall regime and in the presence
of a lateral junction has been carried out in Ref. \cite{kang}.
Our quantitative analysis presented in Fig. 3,4 shows all
important features observed in \cite{kang}, namely a great
enhancement and fast oscillations of the linear conductance $G_x
(0)$ as the magnetic field was decreased (a decrease of the
magnetic field results in an effective increase of the chemical
 potential $\mu$), and unique giant oscillations of $G_x(V)$.

  \begin{figure}
 \centerline{\includegraphics[width=1\columnwidth]{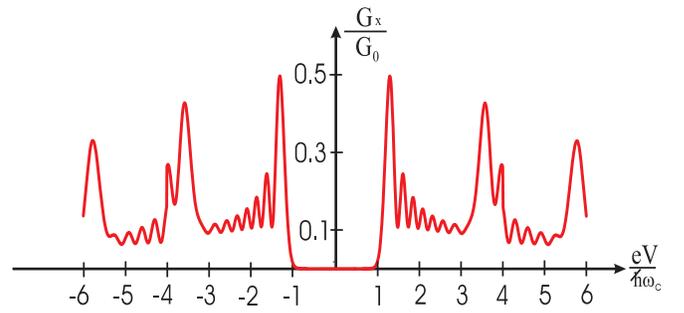}}
 \caption{\textbf{Dependence of the normalized conductance
on the applied dc voltage.}
 The parameters   $\alpha=10^{-2}$,  $\mu_0/\hbar \omega_c=1$, $v/ l_c \omega_c=2$
 and $k_B T/\hbar \omega_c=10^{-2}$ have been used
 (here $G_0= (4 e/h)\gamma^2$ and $\mu_0$ is
 the chemical
 potential in the absence of the applied voltage).}
 \label{conductance}
\end{figure}
In conclusion, we have shown that in the quantum Hall regime a lateral junction formed  in the QPC serves as a unique
quantum-mechanical scatterer for propagating magnetic edge states. Such a lateral junction hosts the electrons bound states having a
peculiar magnetic field dependent 1D spectrum
$\varepsilon_n(p_y)$. This spectrum contains an alternating
sequence of narrow energy bands and gaps. These band and gaps
manifests themselves by  giant oscillations of the transmission
coefficient $T(E)$ on the electron energy $E$ and the voltage
dependent conductance $G_x(V)$. Our theoretical analysis based on the coherent quantum-mechanical superposition of localized and delocalized magnetic edge states  is in a good accord with the experemental observations \cite{kang}. Such a generic  approach [see the Eqs. (\ref{BasicEquation}) and (\ref{probability2})] can be applied to the variety of solid state systems where the physical properties are determined by the coherent quantum dynamics of interacting quantum-mechanical objects, e.g. the magnetic edge states propagation in graphene based nanostructures \cite{grapheneLat} or the superconducting quantum metamaterials \cite{SM}.

\textbf{Acknowledgements:}The financial support of the Ministry of Education and Science of
the Russian Federation  in the framework of
Increase Competitiveness Program of NUST "MISiS"($K2-2014-015$) and
the SPP programme "Graphene" of the DFG are acknowledged.

\section*{Supplementary Information}

\appendix{}

\section{Derivation  of Eq.(5) of
the main text.}

\emph{In this section we find the differential equation  that
describes dynamics of electrons in the vicinity of the points of
degeneration of the left and right edge states travelling along
the lateral barrier placed across the point contact as is shown in
Fig.1 of the main text. }

Inserting $Q$ of Eq.(2) of the main text into Eq.(1)  we obtain
the following equation:

 %
%
\begin{eqnarray}
  \sum_{n=0}^{\infty}\Big\{ R_n \Big[\frac{\partial^2 }{\partial \xi^2}-
  \Big(\frac{p_y \ell_c}{\hbar} -\xi \Big)^2
  - \nu (\xi)\Big] \varphi_{n,p_y}(\xi) \nonumber \\
  +
  \Big[-\left(\frac{\omega_1}{\omega_c}\right)^2 \frac{\hbar^2}{\ell_c^2}
  \frac{d^2 R_n}{d p_y^2}  -2\varepsilon R_n \Big]\varphi_{n,p_y}(\xi)
  \Big\} =0.
\label{SupplRphi}
\end{eqnarray}
%
Here $\varphi_{n,p_y}(\xi)$  are   proper functions of the
following  Schr\"odinger's equation:
%
%
\begin{eqnarray}
  \frac{\partial^2 \varphi_{n,p_y}}{\partial \xi^2}-
  \Big[\Big(\frac{p_y \ell_c}{\hbar} -\xi \Big)^2+
 \nu (\xi)-2 \varepsilon_{n}(p_y) \Big]\varphi_{n,p_y}(\xi) =0,
\label{SupplShcroedinger2}
\end{eqnarray}
%
%
where $\varepsilon_{n}(p_y)$ is the proper energy and $n$ is the
Landau number, the boundary conditions  being
$\varphi_{n,p_y}(\xi) \rightarrow 0$
 at $\xi \rightarrow \pm \infty$.

Eq.(\ref{SupplShcroedinger2})  describes dynamics of electrons
under magnetic field in the presence of a longitudinal  potential
 barrier $\nu (\xi)$ but in the absence of any constriction in the
 direction parallel to the barrier. This situation was considered
 in Refs.\cite{Koshkin,Kang,longitudinal}.  It was shown that
 far from the barrier, $|x| > 2 R_L$ (here $R_L$ is the Larmour radius),
the electron spectrum $\varepsilon_{n}$ is the standard discrete
Landau spectrum. The situation drastically changes for electrons
travelling  along the barrier. In this case the quantum
interference of the edge states situated on its both  sides
transforms the electron spectrum into an alternating series of
narrow energy bands and gaps (see Fig.\ref{trajectories}). As a
result,   both dynamics and kinetics of electrons in such  a
system qualitatively change.

In this section we find the differential equations that describe
evolution of functions $R_n(p_y)$ in the vicinity of the "crossing
points" (e.g., one of them is at $p_y=0$ and $\varepsilon =
\varepsilon_0$, see Fig.\ref{trajectories}) where the quantum
tunnelling of electrons through the lateral junction lifts the
degeneracy in the electron spectrum.
Here we follow  Landau's method for finding energy levels in a
double well potential \cite{Landau}.
%


In the limit of a low barrier transparency the wave functions
$\varphi_{n,p_y}(\xi)$ and $\varphi_{n+1,p_y}(\xi)$ close to the
"crossing points"
 are conveniently presented as a combination of the left
 $\phi_{n,p_y}^{(l)}(\xi)$ and right $\phi_{n^\prime,p_y}^{(r)}(\xi)$
independent edge states  the proper energies of which intersect in
the middle of the gap.   These states are localized in the left
well, $\xi <0$, and the right one,
 $\xi >0$, respectively, and hence one may write:
%
\begin{eqnarray}
 \varphi_{n,p_y}(\xi)\approx \frac{1}{\sqrt{2}}\big( \phi_{n,1}(\xi)
  +\phi_{n,2}(\xi)\big)\nonumber \\
\varphi_{n+1,p_y}(\xi)\approx \frac{1}{\sqrt{2}}\big(
\phi_{n,1}(\xi)
  -\phi_{n,2}(\xi)\big);
\label{Suppledge0}
\end{eqnarray}
%
%
where $\phi_{n,1}(\xi)\equiv \phi_{n,p_y}^{(l)}(\xi)$ and
$\phi_{n,2}(\xi)\equiv \phi_{n^\prime,p_y}^{(r)}(\xi)$ are the
edge state wave functions which satisfy the following equations:
%
%
\begin{eqnarray}
  \Big\{\frac{\partial^2 }{\partial\xi^2}
- \Big(\frac{p_y \ell_c}{\hbar} -\xi \Big)^2 -
 \nu (\xi)+2 \overline{\varepsilon}_{1,2}^{(n)}(p_y) \Big\}\phi_{n,1,2}(\xi)=0
\label{SuplShcroedingerEdge}
\end{eqnarray}
%
%
with the boundary conditions  $\phi_{1}^{(n)}(\xi) \rightarrow 0$
 on the both boundaries of
 the left well  while  $\phi_{2}^{(n)}(\xi) \rightarrow 0$   on the both
 boundaries of the right
 one. Here $\overline{\varepsilon}_{1,2}^{(n)}(p_y)$ are the energies of edge states
 (see, e.g. \cite{Dittrich})
 which degenerate at $p_y =p_y^{n}$.
Four such points at $p_y=0, \, p_y=p^{(I,II)}$, are shown in
Fig.\ref{trajectories}.
%
%
%

 Multiplying Eq.(\ref{SupplRphi}) by $\phi_{n,1}(\xi)$  and
 integrating over $p_y$ from $-\infty$ to $a$ as well as by
 $\phi_{n,2}(\xi)$ and integrating from $a$ to $+\infty$ one
 gets the following set of coupled  equations
%
%
\begin{eqnarray}
\sum_{\bar{n}=0}^{\infty} \left[\phi_{n,1}(a)
\varphi_{\bar{n},p_y}^{\prime}(a)
  -\phi_{n,1}^{\prime}(a)
  \varphi_{\bar{n},p_y}(a)  \right]R_{\bar{n}}(p_y)  \nonumber \\
 +\Big[ \overline{\varepsilon}_1^{(n)}(p_y) -\varepsilon -
 \frac{\hbar^2}{2\ell_c^2}\alpha^2 \frac{d^2}{d p_y^2}\Big]
 \nonumber \\
\times \sum_{\bar{n}=0}^\infty R_{\bar{n}}\int_{-\infty}^a
 \phi_{n,1}(\xi)  \varphi_{n,p_y}(\xi)d\xi=0;\nonumber \\
\sum_{\bar{n}=0}^\infty  \left[-\phi_{n,2}(a)
\varphi_{\bar{n},p_y}^{\prime}(a)
  +\phi_{n,2}^{\prime}(a)
  \varphi_{\bar{n},p_y}(a)  \right]R_{\bar{n}}(p_y)  \nonumber \\
 +\Big[ \overline{\varepsilon}_2^{(n)}(p_y) -\varepsilon -
 \frac{\hbar^2}{2\ell_c^2}\alpha^2 \frac{d^2}{d p_y^2}\Big]\nonumber \\
\times
 \sum_{\bar{n}=0}^\infty  R_{\bar{n}}\int_a^{\infty}
 \phi_{n,2}(\xi)  \varphi_{\bar{n},p_y}(\xi)d\xi=0;\
\label{GF2}
\end{eqnarray}
%
%
Here $a$  is the point inside of the barrier at which
$\varphi_{n,p_y}^{\prime}(a) =0$ and $\alpha
 =\omega_1/\omega_c$ while $f^{\prime}(\xi) \equiv d f (\xi)/d \xi$.

As the edge state functions are orthogonal and normalized while
$\varphi_{n,p_y}(\xi)$ are approximated by Eq.(\ref{Suppledge0})
one finds
%
%
\begin{eqnarray}
\sum_{\bar{n}=0}^\infty R_{\bar{n}} \int_{-\infty}^a
 \phi_{n,1}(\xi)  \varphi_{\bar{n},p_y}(\xi)d\xi=\frac{1}{\sqrt{2}}\Big(R_n(p_y)
 +R_{n+1}(p_y) \Big);\nonumber \\
 \sum_{\bar{n}=0}^\infty R_{\bar{n}}\int_a^{+\infty}
 \phi_{n,2}(\xi)  \phi_{\bar{n},p_y}(\xi)d\xi=\frac{1}{\sqrt{2}}\Big(R_n(p_y)
 -R_{n+1}(p_y) \Big);
 \label{SumRsimple1}
\end{eqnarray}
%
%
and
%
%
\begin{eqnarray}
\sum_{\bar{n}=0}^{\infty} \left[\phi_{n,1,2}(a)
\varphi_{\bar{n},p_y}^{\prime}(a)
  -\phi_{n,1,2}^{\prime}(a)
  \varphi_{\bar{n},p_y}(a)  \right]R_{\bar{n}}(p_y)\approx  \nonumber \\
 -\phi_{n,1,2}^\prime(a)
\frac{\phi_{n,1}(a)+\phi_{n,2}(a)}{\sqrt{2}}\Big(R_n(p_y)\mp
R_{n+1}(p_y)\Big)
 \label{SumRsimple2}
\end{eqnarray}
%
%
Inserting Eqs.(\ref{SumRsimple1},\ref{SumRsimple2}) in
Eq.(\ref{GF2}) one gets the set of differential equations
presented in the main text (Eq.(5)):

%
%
\begin{eqnarray}
\left(\alpha^2\frac{\hbar^2}{2\ell_c^2}  \frac{d^2 }{d p_{y}^2} +
+\overline{\varepsilon}_n^{(l)}(p_y)- \varepsilon \right)
R_{n}^{(l)} +
\tilde{\Delta}_n R_{n}^{(r)} =0;\nonumber \\
\left(\alpha^2\frac{\hbar^2}{2\ell_c^2}  \frac{d^2 }{d p_y^2} +
\overline{\varepsilon}_n^{(r)}(p_y)- \varepsilon
 \right)
 R_{n}^{(r)} +\tilde{\Delta}_n R_{n}^{(l)} =0,
\label{BasicEquation}
\end{eqnarray}
%
%
Here $R_{n}^{(l,r)}=R_n \mp R_{n+1} $, and
%
%
\begin{eqnarray}
\tilde{\Delta}_n = -\phi_{n,1}^\prime(a)
\frac{\phi_{n,1}(a)+\phi_{n,2}(a)}{\sqrt{2}}
 \label{gap}
\end{eqnarray}
%
%
is the normalized energy gap in the spectrum of the bound states.
For the sake of simplicity, below we drop the subscript $n$
assuming that all energy gaps are equal.

\emph{\textbf{Using Eq.(\ref{BasicEquation}) at $n=0$ one obtains
Eq.(5) of the main text.} }

\section{Derivation of the probability $D(E)$ of the electron transmission
from $p_y \rightarrow -\infty$  to $p_y \rightarrow +\infty$,
Eq.(11) of the main text.}

\emph{1 step. In  subsection A  we find the semiclassical
solutions of Eq.(\ref{BasicEquation}) and we show that  the
semiclassical approximation fails not only at the turning points
but in narrow vicinities of the degenerate points as well where
the interband quantum transitions between semiclassical
trajectories  (see Fig. 2B) takes place. }

\emph{ 2 step. Therefore, before finding the probability for an
electron to pass from $p_y = -\infty$ to $p_y=+\infty$,   we solve
Eq.(\ref{BasicEquation}) in the vicinity of the degeneration
points developing some perturbation theory, see subsection B. We
also show that there are regions on the both sides of the
degeneration points where this solution overlaps with the
semiclassical wave functions found in subsection A. The found
solution allows to re-write it in terms of the probability
transitions between these regions of the overlap, this probability
$D(E)$ being the same as presented in Eq.(11) of the main text.}

\emph{3 step. In subsection C,  in order to prove that $D(E)$ is
also the probability for the electron to pass from $p_y = -\infty$
to $p_y = +\infty$ we match the semiclassical wave functions and
the perturbation ones in the regions of overlapping, and then we
take asymptotic of the former functions at $p_y = \pm \infty$ and
find Eq.(11) of the main text. }
\subsection{Quasiclassical wave function in the momentum space. }

Here we find  quasiclassical solutions of Eq.(\ref{BasicEquation})
using the inequality $\alpha \ll 1$.

Presenting functions $R_n^{(l,r)}$ in the quasiclassical form
%
%
\begin{eqnarray}
R_n^{(l,r)}(p_y)=
A_n^{(l,r)}(p_y)\exp\{i\frac{S_n(p_y)\ell_c}{\hbar \alpha} \}
 \label{SupplSemiclassic}
\end{eqnarray}
%
%
one easily finds the classical momentum $P_n= d S_n/dp_y$, the
semiclassical parameter $\kappa$
%
%
\begin{eqnarray}
P_n^2=\varepsilon - \frac{\overline{\varepsilon}_n^{(l)}(p_y) +
\overline{\varepsilon}_n^{(r )}(p_y)}{2}\nonumber \\
 \pm
\sqrt{\Big(\frac{ \overline{\varepsilon}_n^{( l)}(p_y) -
\overline{\varepsilon}_n^{( r)}(p_y)}{2}\Big)^2+\tilde{\Delta}^2}; \nonumber \\
\kappa =(\frac{\omega_c \ell_c}{\alpha v})^{2/3}|\varepsilon -
\overline{\varepsilon}_n^{(l,r )}(p_y)| \gg 1
\label{SupplMomentum}
\end{eqnarray}
%
%
 and the wave
functions:
\begin{eqnarray}
R_n^{(l)}(p_y)(p_y)=\frac{1}{(\varepsilon
-\overline{\varepsilon}_n^{(l )})^{1/4}}
\nonumber \\
\times
\Big[B_{1}^{(n)}\exp\left(\frac{i\ell_c}{\hbar\alpha}\int^{p_y}P_1^{(n)}
d
p^\prime_y\right)\nonumber \\
+ B_{2}^{(n)}
\exp\left(-\frac{i\ell_c}{\hbar \alpha}\int^{p_y}P_1^{(n)} d p^\prime_y\right)\Big]\nonumber \\
+\frac{\tilde{\Delta}}{\overline{\varepsilon}_n^{(r )}
-\overline{\varepsilon}_n^{(l)}}
\times \frac{1}{(\varepsilon -\overline{\varepsilon}_n^{(r )})^{1/4}}\nonumber \\
\times
\Big[C_{1}^{(n)}\exp\left(\frac{i\ell_c}{\hbar\alpha}\int^{p_y}P_2^{(n)}
d
p^\prime_y\right)\nonumber \\
+
C_{2}^{(n)}\exp\left(-\frac{i\ell_c}{\hbar\alpha}\int^{p_y}P_2^{(n)}
d p^\prime_y\right)\Big] \label{SupplSemiclassic1}
\end{eqnarray}
%
%
%
%
\begin{eqnarray}
R_n^{(r)}(p_y)(p_y)=-\frac{\tilde{\Delta}}{\overline{\varepsilon}_n^{(r)}
-\overline{\varepsilon}_n^{(l)}}\times\frac{1}{(\varepsilon
-\overline{\varepsilon}_1^{(n)})^{1/4}}
\nonumber \\
\times \Big[B_{1}^{(n)}\exp\left(\frac{i\ell_c}{\alpha
\hbar}\int^{p_y}P_1^{(n)} d
p^\prime_y\right)\nonumber \\
+ \Big(B_{2}^{(n)}
\exp\left(-\frac{i\ell_c}{\alpha \hbar}\int^{p_y}P_1^{(n)} d p^\prime_y\right)\Big]\nonumber \\
+\frac{1}{(\varepsilon -\overline{\varepsilon}_n^{(r )})^{1/4}}\nonumber \\
\times \Big[C_{1}^{(n)}\exp\left(\frac{i\ell_c}{\alpha
\hbar}\int^{p_y}P_2^{(n)} d
p^\prime_y\right)\nonumber \\
+ \Big(C_{2}^{(n)} \exp\left(-\frac{i\ell_c}{\alpha
\hbar}\int^{p_y}P_2^{(n)} d p^\prime_y\right)\Big]
\label{SupplSemiclassic2}
\end{eqnarray}
%
%
where  $B_{1,2}^{(n)}$ and $C_{1,2}^{(n)}$ are arbitrary
constants, $P_{1,2}^{(n)}$ are two solutions  of
Eq.(\ref{SupplMomentum}) corresponding to $\pm$ at the squire
root.

Quasiclassical trajectories $P_n=P_n(p_y; \varepsilon)$
corresponding to Eq.(\ref{SupplMomentum}) for two electron
energies above the first energy gap (the Landau number $n=0$) and
above the next two ones (the Landau number $n=1$) are shown in
Fig.\ref{trajectories}b.
  \begin{figure}
 \centerline{\includegraphics[width=0.7\columnwidth]{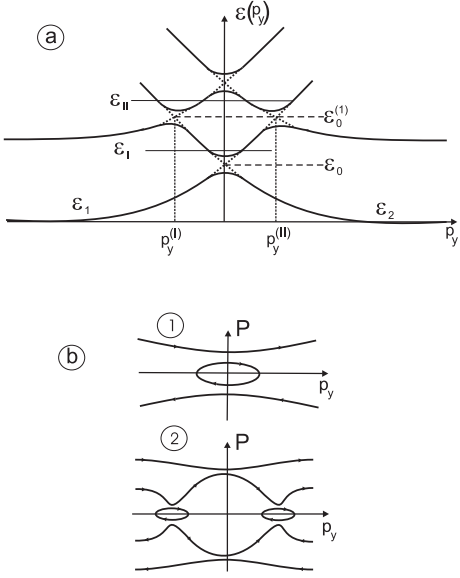}}
 \caption{a) Energy spectrum of the quasiparticles localized around the longitudinal
  barrier; bI and bII - semiclassical trajectories $P=P(p_y;\varepsilon)$ corresponding to
  $\varepsilon=\varepsilon_I$ and $\varepsilon=\varepsilon_{II}$, respectively.}
 \label{trajectories}
 \end{figure}

As one sees from
Eqs.(\ref{SupplSemiclassic1},\ref{SupplSemiclassic2}) the
quasiclassical approximation fails not only at the turning point
but at any degeneration point $p_y^{(n)}$ at which
$\overline{\varepsilon}_n^{(l )} (p_y) =
\overline{\varepsilon}_n^{(r )} (p_y)$ . In the next section, in
order to match the semiclassical wave functions  we solve
Eq.(\ref{BasicEquation}) in the vicinity of the crossing points by
making use of the perturbation analysis .

\subsection{Derivation of the transfer matrix  in the vicinity of a degeneration point. }
We start derivation of the transfer matrix  solving
Eq.(\ref{BasicEquation})  in the immediate vicinity of a
degenerate point.  Expanding the edge state energies
$\overline{\varepsilon}_n^{(l,r )} (p_y)\approx
\varepsilon_0^{(n)} \mp v_n^{(l,r )} \widetilde{p}_y$ (for the
sake of certainty these signs of the edge state velocities are
chosen  to correspond to
 degeneration points $p_y^{(n)} \leq 0$, see
Fig.\ref{trajectories}, $\widetilde{p}_y=p_y-p_y^{(n)}$) one
obtains the following
  equations:
%
%
\begin{eqnarray}
\left(\alpha^2 \frac{\hbar^2}{2\ell_c^2} \frac{d^2 }{d
\widetilde{p}_{y}^2} +\varepsilon -\varepsilon_0^{(n)} -v_n^{(l)}
\widetilde{p}_y \right) R_n^{(l )} -
\tilde{\Delta} R_n^{(r )} =0;\nonumber \\
\left(\alpha^2\frac{\hbar^2}{2\ell_c^2}  \frac{d^2 }{d
\widetilde{p}_y^2} +\varepsilon -\varepsilon_0^{(n)} +v_n^{(r )}
\widetilde{p}_y\right)
 R_n^{(r )} -\tilde{\Delta} R_n^{(l )} =0,
\label{BasicReduced}
\end{eqnarray}
%
%
where $v_n^{(l,r )} =|d\overline{ \varepsilon}_n^{(l,r )}/dp_y|$
taken at $p_y=p_y^{(n)}$.

 To reduce the order of the differential
equations we perform the Fourier transformation as
%
%
\begin{eqnarray}
R_n^{(l,r )}(p_y)= \int_{-\infty}^{+\infty}
g_{1,2}^{(n)}(Y)\exp{\left(\frac{i\ell_c}{\hbar\alpha}Y
\widetilde{p}_y\right)}dY
 \label{Fourier3}
\end{eqnarray}
%
%
and get the following set of equations:
%
%
\begin{eqnarray}
i \alpha v_1^{(n)} \frac{d g_1^{(n)} }{d  Y} -(\varepsilon
-\varepsilon_0^{(n)} -Y^2 ) g_1^{(n)} +
\tilde{\Delta} g_2^{(n)} =0;\nonumber \\
i \alpha v_2^{(n)} \frac{d g_2 }{d  Y} +(\varepsilon
-\varepsilon_0^{(n)} -Y^2 ) g_2^{(n)} - \tilde{\Delta} g_1^{(n)}
=0; \label{BasicReducedG}
\end{eqnarray}
\subsubsection{Transfer matrix in the $Y$-space.}
Using Eq.(\ref{BasicReducedG}) and presenting  the Fourier factors
$g_{1,2}^{(n)}(Y)$ in the form
%
%
\begin{eqnarray}\label{g}
g_{1}^{(n)}=u_{1}^{(n)}(\zeta)\exp\{-i\left(\frac{v_2^{(n)}}{v_1^{(n)}+v_2^{(n)}}(\zeta^3/3+
\eta_n\zeta)\right)\}\nonumber \\
g_{2}^{(n)}=u_{2}^{(n)}(\zeta)\exp\{+i\left(\frac{v_1^{(n)}}{v_1^{(n)}+v_2^{(n)}}(\zeta^3/3+
\eta_n\zeta)\right)\}
\end{eqnarray}
%
%
together with a  change of the variable  $$\zeta=Y[\frac{\ell_c
\omega_c}{\alpha \overline{v}_n)}]^{1/3};
\;\;\;\overline{v}_n=\frac{v_1^{(n)}v_2^{(n)}}{v_1^{(n)}+v_2^{(n)}}$$
 one gets the following
parameterized set of equations
%
%
\begin{eqnarray}\label{SupplSystemH3}
 i\frac{du_1^{(n)}(\zeta)}{d\zeta}=+\overline{\gamma}_n
\frac{v_2^{(n)}}{v_1^{(n)}+v_2^{(n)}}
e^{+i\left(\frac{2}{3}\zeta^3 +
\eta_n \zeta\right)} u_2(\zeta) \nonumber \\
 i\frac{du_1^{(n)}(\zeta)}{d\zeta}=-\overline{\gamma}_n
\frac{v_1^{(n)}}{v_1^{(n)}+v_2^{(n)}}
e^{-i\left(\frac{2}{3}\zeta^3 + \eta_n \zeta\right)} u_2(\zeta) ,
\end{eqnarray}
where
%
%
\begin{eqnarray}
\overline{\gamma}_n =  \tilde{\Delta}[\frac{\ell_c
\omega_c}{\alpha \overline{v}_n}]^{2/3}; \hspace{0.2cm} \eta_n
=(\varepsilon_0^{(n)} -\varepsilon)[\frac{\ell_c \omega_c}{\alpha
\overline{v}_n} ]^{2/3}.
 \label{Parameters}
\end{eqnarray}
%
%
We note here that analogous set of equations (with
$v_1^{(n)}=v_2^{(n)}=v_F$) define the magnetic breakdown
probability near  touching points of classical orbits of an
electron under a strong magnetic field as it was shown and
analyzed in paper \cite{PRL}.

We assume $\gamma_n \ll 1$ and find  the following solution of
Eq.(\ref{SupplSystemH3}) by the perturbation theory:
%
%
\begin{eqnarray}
u_1^{(n)}(\zeta)=-i\overline{\gamma}_n
\frac{v_2^{(n)}}{v_1^{(n)}+v_2^{(n)}}
u_{2}^{(n)}(-\infty)\int_{-\infty}^{\zeta}d \chi
 e^{if(\chi)} + \nonumber \\
u_{1}^{(n)}(-\infty)\Big[1+ (\overline{\gamma}_n
\overline{w}_n)^2\int_{-\infty}^{\zeta} d \chi e^{if(\chi)}
\int_{-\infty}^{\chi}d
\chi^{\prime} e^{-if(\chi^{\prime})}\Big]\nonumber \\
u_2^{(n)}(\zeta)=i\overline{\gamma}_n
\frac{v_2^{(n)}}{v_1^{(n)}+v_1^{(n)}}
u_{1}^{(n)}(-\infty)\int_{-\infty}^{\zeta}d \chi
 e^{-if(\chi)}  + \nonumber \\
u_{2}^{(n)}(-\infty)  \Big[1+ (\overline{\gamma}_n
\overline{w}_n)^2\int_{-\infty}^{\zeta}d \chi e^{-if(\chi)}
\int_{-\infty}^{\chi}d \chi^{\prime}
e^{if(\chi^{\prime})}\Big]\hspace{0.1cm} \label{SupplU}
\end{eqnarray}
%
%
where
%
%
\begin{eqnarray}
f(\zeta)=\frac{\zeta^3 }{3}+\eta_n \zeta;\nonumber \\
\overline{w}_n
=\frac{\sqrt{v_1^{(n)}v_2^{(n)}}}{v_1^{(n)}+v_2^{(n)}};
 \label{fw}
\end{eqnarray}
%
%

Taking   $|\zeta| \gg 1 $ one gets the transfer matrix
\begin{eqnarray}
\hat{\tau_n} =\left( \begin{array}{ll} 1+(\overline{\gamma}_n
\overline{w}_n)^2 a_2^{(n)}&i \overline{\gamma}_n
\frac{v_2^{(n)}}{v_1^{(n)}+v_2^{(n)}}
a_1^{(n)}\\
-i \overline{\gamma}_n \frac{v_1^{(n)}}{v_1^{(n)}+v_2^{(n)}}
a_1^{(n) \star}&1+(\overline{\gamma}_n \overline{w}_n)^2 a_2^{(n)
\star}
\end{array}\right)
\label{transfer0}
\end{eqnarray}
that couples constants $u_{1,2}^{(n)}(-\infty)$ in the solutions
of Eq.(\ref{SupplSystemH3}) at $\zeta < 0, -\zeta \gg 1$ and their
asymptotic  $u_{1,2}^{(n)}(+ \infty)$ at $\zeta >0, \zeta \gg 1$
as
\begin{eqnarray}
\left(\begin{array}{ll} u_1^{(n)}(+\infty) \\
u_2^{(n)}(+\infty) \end{array} \right) = \hat{\tau} \left(\begin{array}{ll} u_1^{(n)}(-\infty)\\
u_2^{(n)}(-\infty)
\end{array}\right)
\label{transfer1}
\end{eqnarray}
Here,
%
\begin{eqnarray}
a_1^{(n)}\equiv a_1 (\eta_n)= \int_{-\infty}^{+\infty}
\exp{\left\{-i \big(
\frac{\chi^3}{3} +\eta_n \chi\big) \right\}}d \chi \nonumber \\
=2 \pi \textrm{Ai}\left(\eta_n\right); \label{a1}
\end{eqnarray}
and
%
\begin{eqnarray}
a_2^{(n)}=\int_{-\infty}^{+\infty} d \chi \exp{\left\{-i \big(
\frac{\chi^3}{3} +\eta_n \chi\big) \right\}}\nonumber \\
\times \int_{-\infty}^{\chi} d \tau \exp{\left\{+i \big(
\frac{\chi^3}{3} +\eta_n\tau\big) \right\}}, \label{a2}
\end{eqnarray}
where $\textrm{Ai}\left(x\right)$ is the Airy function; one easily
sees that $a_2 + a_2^{\star}=|a_1|^2$.

 In the next subsection we calculate   the asymptotic of
Eq.(\ref{Fourier3}) at large $|\widetilde{p_y}|$ to use it later
for matching the quasiclassical wave functions
Eqs.(\ref{SupplSemiclassic1},\ref{SupplSemiclassic2}) on the both
sides of the degeneration point $p_y^{(n)}$.

\subsubsection{Derivation of the transmission matrix that couples
the semiclassical wave functions in the vicinity of the
degeneration point.}

Using Eqs.(\ref{Fourier3},\ref{g}) one finds the wave functions in
the momentum space as follows:
%
%
\begin{eqnarray}
R_n^{(l)}(\overline{p}_y)= (\alpha \overline{v}_n)^{1/3}
\int_{-\infty}^{+\infty} u_1^{(n)}(\zeta) \exp\Big\{-i
\Big[\frac{v_2^{(n)}}{v_1^{(n)}+v_2^{(n)}}\frac{\zeta^3 }{3}\nonumber \\
+\Big(\frac{v_2^{(n)}}{v_1^{(n)}+v_2^{(n)}}\eta_n
-\frac{1}{\hbar}(\frac{\overline{v}_n \ell_c^2}{\alpha^2
\omega_c})^{1/3}\widetilde{p_y}\Big)\zeta
\Big]\Big\}d\zeta \nonumber \\
R_n^{(r )}(\overline{p}_y)= (\alpha \overline{v}_n)^{1/3}
\int_{-\infty}^{+\infty} u_2^{(n)}(\zeta) \exp\Big\{+i
\Big[\frac{v_2^{(n)}}{v_1^{(n)}+v_2^{(n)}}\frac{\zeta^3 }{3}\nonumber \\
+\Big(\frac{v_2^{(n)}}{v_1^{(n)}+v_2^{(n)}}\eta_n
+\frac{1}{\hbar}(\frac{\overline{v}_n \ell_c^2}{\alpha^2
\omega_c})^{1/3}\widetilde{p_y}\Big)\zeta \Big]\Big\}d\zeta
\hspace{0.5cm}
 \label{AsymptoticPhi}
\end{eqnarray}
%
%
We perform the integration over $\zeta$ by the steepest descent
method at $|p_y -p_y^{(n)}| \gg (\alpha^2 \ell_c
\omega_c/\overline{v}_n)^{1/3} (\hbar/\ell_c)$.

As one sees from Eq.(\ref{AsymptoticPhi}) the saddle points are
%
%
\begin{eqnarray}
\zeta_{\pm}^{(1)}= \pm \frac{(\ell_c \omega_c)^{1/3}}{(\alpha
\overline{v}_n)^{1/3}}\sqrt{\varepsilon - \varepsilon_0^{(n)}+
v_n^{(l)}p_y};
 \label{saddle1}
\end{eqnarray}
%
%
for the integral in the expression for $R_n^{(l)}$ and
%
%
\begin{eqnarray}
\zeta_{\pm}^{(2)}= \pm \frac{(\ell_c \omega_c)^{1/3}}{(\alpha
\overline{v}_n)^{1/3}}\sqrt{\varepsilon - \varepsilon_0^{(n)}-
v_n^{(r )}p_y};
 \label{saddle2}
\end{eqnarray}
%
%
for the integral in the expression for $R_n^{(r )}$.

Using  Eqs.(\ref{AsymptoticPhi}-\ref{saddle2}) one finds the
asymptotic of $ R_n^{(l,r )}$ at $p_F \gg |p_y-p^{(n)}_y|  \gg
(\alpha^2/ v)^{1/3}$  (here, $v_F$ and $p_F$ are
 the Fermi velocity and momentum) as follows:
%
%
\begin{eqnarray}
p_y -p_y^{(n)}<0, \hspace{5cm} \nonumber \\
R_n^{(l )}(p_y)  \propto \exp{\Big[-\frac{2}{3} \frac{\ell_c
\omega_c
\big(\widetilde{\varepsilon}_n^{(l)}(p_y)-\varepsilon\big)^{3/2}}{\alpha
v_n^{(l)}}\Big]};  \hspace{0.5cm} \nonumber \\
 R_n^{(r )}(p_y)= \frac{\sqrt{\pi \alpha
 v_n^{(r )}}}{\big(\varepsilon-\widetilde{\varepsilon}_n^{(r )}(p_y)\big)^{1/4}} \hspace{2.5cm} \nonumber \\
\times \Big\{u_2^{(n)}(-\infty)\exp{\Big[i\Big(\frac{2}{3}
\frac{\ell_c \omega_c
\big(\varepsilon-\widetilde{\varepsilon}_n^{(r)}(p_y)\big)^{3/2}}{\alpha
v_n^{(r )}}+\frac{\pi}{4}\Big) \Big]} \nonumber \\
+u_2^{(n)}(+\infty)\exp{\Big[-i\Big(\frac{2}{3} \frac{\ell_c
\omega_c\big(\varepsilon-\widetilde{\varepsilon}_n^{(r)}(p_y)\big)^{3/2}}{\alpha
v_n^{(r )}}+\frac{\pi}{4}\Big) \Big]}\Big\};
 \label{AsymptoticPhi1}
\end{eqnarray}
%
%
%
%
\begin{eqnarray}
p_y -p_y^{(n)}>0, \hspace{5cm} \nonumber \\
R_n^{(l)}(p_y)= \frac{\sqrt{\pi \alpha
 v_n^{(l )}}}{\big(\varepsilon-\widetilde{\varepsilon}_n^{(l )}(p_y)\big)^{1/4}}
 \hspace{2.5cm} \nonumber \\
\times \Big\{u_1^{(n)}(-\infty)\exp{\Big[-i\Big(\frac{2}{3}
\frac{\ell_c
\omega_c\big(\varepsilon-\widetilde{\varepsilon}_n^{(l)}(p_y)\big)^{3/2}}{\alpha
v_n^{(l )}}-\frac{\pi}{4}\Big) \Big]} \nonumber \\
+u_1^{(n)}(+\infty)\exp{\Big[+i\Big(\frac{2}{3} \frac{\ell_c
\omega_c\big(\varepsilon-\widetilde{\varepsilon}_n^{(l)}(p_y)\big)^{3/2}}{\alpha
v_n^{(l )}}-\frac{\pi}{4}\Big) \Big]}\Big\}; \nonumber \\
R_n^{(r )}(p_y)  \propto \exp{\Big[-\frac{2}{3} \frac{\ell_c
\omega_c\big(\widetilde{\varepsilon}_n^{(r)}(p_y)-\varepsilon\big)^{3/2}}{\alpha
v_n^{(r )}}\Big]}, \hspace{0.5cm} \label{AsymptoticPhi2}
\end{eqnarray}
%
%
where $\widetilde{\varepsilon}_n^{(l,r )}(p_y)=
\varepsilon_0^{(n)} \mp v_n^{(l,r )}(p_y-p_y^{(n)})$ are the edge
state energies expanded near the degenerate point $p_y^{(n)}$.

Matching  the wave functions Eqs.(\ref{AsymptoticPhi1},
\ref{AsymptoticPhi2}) and
 the quasiclassical wave functions Eqs.(\ref{SupplSemiclassic1},\ref{SupplSemiclassic2})
 in the overlapping
 region $p_F \gg |p_y-p_y^{(n)}| \gg (\ell_c \omega_c\alpha^2/ v_F)^{1/3} (\hbar/\ell_c)$
 we obtain
%
%
\begin{eqnarray}
p_y-p_y^{(n)}< 0 \hspace{5cm} \nonumber \\
C_1^{(n)}=\sqrt{\alpha \pi v_n^{(r )}}e^{-i\pi/4} u_2^{(n)}(+\infty);\nonumber \\
C_2^{(n)}=\sqrt{\alpha \pi v_n^{(r )}}e^{+i\pi/4} u_2^{(n)}(-\infty);\nonumber \\
p_y-p_y^{(n)}> 0 \hspace{5cm} \nonumber \\
B_1^{(n)}=\sqrt{\alpha \pi v_n^{(l )}}e^{-i\pi/4} u_1^{(n)}(+\infty);\nonumber \\
B_2^{(n)}=\sqrt{\alpha \pi v_n^{(l)}}e^{+i\pi/4}
u_1^{(n)}(-\infty); \label{constants}
\end{eqnarray}
%
The other constants one may put equal to zero.

Using Eq.(\ref{constants}) and
Eqs.(\ref{transfer0},\ref{transfer1}) one finds the transmission
matrix that couples the amplitudes of the incoming,
$B_2^{(n)},\;C_1^{(n)}$, and outgoing, $B_1^{(n)},\;C_2^{(n)}$,
quasiclassical wave functions
Eqs.(\ref{SupplSemiclassic1},\ref{SupplSemiclassic2}) on the both
sides of the region around $p_y^{(n)}$
\begin{eqnarray}
\left(\begin{array}{ll}
B_1^{(n)} \\
C_2^{(n)} \end{array} \right) = \hat{t}_n \left(\begin{array}{ll} B_2^{(n)}\\
C_1^{(n)}
\end{array}\right)
\label{transmission1}
\end{eqnarray}
where
\begin{eqnarray}
\hat{t}_n = e^{i \Theta}\left( \begin{array}{ll}
-1+\frac{(\overline{\gamma}_n
\overline{w}_n)^2}{2}|a_1^{(n)}|^2&\overline{\gamma}_n
\overline{w}_n a_1^{(n)}\\
\overline{\gamma}_n \overline{w}_n a_1^{(n)}&
1-\frac{(\overline{\gamma}_n \overline{w}_n)^2}{2}|a_1^{(n)}|^2
\end{array}\right)
\label{transmission2}
\end{eqnarray}
and  $\Theta =(\overline{\gamma}_n \overline{w}_n)^2\, Im\{
a_2^{(n)}\}-\pi/2$. Equation Eq.(\ref{transmission2}) is written
with the accuracy of the second order of $\overline{\gamma}_n$
while parameters $\overline{\gamma}_n, \, \overline{w}_n$ are
defined in Eq.(\ref{Parameters}).

The transmission matrix $\hat{t}_n $  couples  the quasiclassical
wave functions
Eqs.(\ref{SupplSemiclassic1},\ref{SupplSemiclassic2}) on the both
sides of any point of intersection of the two  edge state
dispersion laws, e.g., for the case that the energy is in the
vicinity of $\varepsilon_0^{(1)}$ (see Fig.\ref{trajectories})
these regions are $ -\infty \leq p_y \lesssim
p_y^{(I)}-(\alpha^2/\overline{v}_F)^{1/3},
 \; p_y^{(I)}+(\alpha^2/\overline{v}_F)^{1/3} \lesssim p_y
  \lesssim p_y^{(II)}-(\alpha^2/\overline{v}_F)^{1/3}$.

In the next section we the  find electron
 wave functions at $x \rightarrow -\infty$ and $x \rightarrow +\infty$ that
 allows to find the probability of electron transmission through the point
 contact under considerations.

\subsection{The matrix of transmission through the point contact from
$p_y \rightarrow -\infty$  to $p_y \rightarrow +\infty$ .}

To find the probability for an electron to pass through the point
contact one firstly needs to know connections between the
coefficients  in the incoming  and outgoing  quasiclassical wave
functions at $p_y \rightarrow \pm \infty$. As it follows from
Eqs.(\ref{SupplSemiclassic1},\ref{SupplSemiclassic2}), in these
regions the functions are
%
%
\begin{eqnarray}
p_y \rightarrow - \infty;  \hspace{0.5cm}
R_n^{(l)}(p_y)=\frac{1}{\sqrt{p_x^{(n)}}} \hspace{2cm}  \nonumber \\
\times \Big[C_{in}^{(l,n)}
\exp{\left(i\frac{p_x^{(n)}p_y}{p_1^2}\right)}
+ C_{out}^{(l,n)}\exp{\left(-i\frac{p_x^{(n)}p_y}{p_1^2}\right)}\Big] \nonumber \\
p_y \rightarrow + \infty;  \hspace{0.5cm}
R_n^{(r )}(p_y)=\frac{1}{\sqrt{p_x^{(n)}}} \hspace{2cm}  \nonumber \\
\times \Big[C_{out}^{(r,n)}
\exp{\left(i\frac{p_x^{(n)}p_y}{p_1^2}\right)} +
C_{in}^{(r,n)}\exp{\left(-i\frac{p_x^{(n)}p_y}{p_1^2}\right)}\Big]
\label{FuctionsPInfinity}
\end{eqnarray}
%
%
where  $p_x^{(n)}=\sqrt{2 m (E-\hbar \omega_H(n+1/2)}$ , and
$p_1^2=m \hbar \omega_1$; $C_{in}^{(n)}, \; C_{in}^{(n)}$ and
$C_{out}^{(n)}, \; B_{out}^{(n)}$  are the amplitudes of the
incoming and outgoing wave functions normalized to the unity flux.
While writing the above equation we took into account that in the
limiting regions the electron energy $E_n(p_y) \rightarrow \hbar
\omega_H(n+1/2)$.

We start from the case that the electron energy is in the vicinity
of $\varepsilon_0^{(1)}$ as is shown in Fig.\ref{trajectories} and
hence there are two points $p_y^{(I)}$ and $p_y^{(II)}$ at which
the quasiclassical trajectories undergo topological transitions
under a change of the electron energy. The quasiclassical
trajectories at the electron energy above this point, $\varepsilon
>\varepsilon_0^{(1)}$, are presented in Fig.\ref{trajectories}b2.

Matching quasiclassical  wave functions
Eq.(\ref{FuctionsPInfinity}) with the help od the transfer matrix
Eq.(\ref{transmission1},\ref{transmission2}), and taking into
account the quasiclassical phase gains between the degeneracy
points $p_y^{(I,II)}$ one finds the coupling between the outgoing
coefficients $C_{out}^{(r,0)}$, $C_{out}^{(r,1)}$ and the incoming
ones $C_{in}^{(l,0)}$, $C_{in}^{(r,1)}$  as follows:
%
%
\begin{eqnarray}
C_{out}^{(r,n-1)}=t_L\frac{1-|r_{n}|^2\exp\{i
\chi_+\}}{1-|t_{L}|^2|r_{n}|^2\exp\{i \chi_+\}}
C_{in}^{(l,n-1)}\nonumber \\
 -r_L \frac{t_n^{\star}\exp\{i \chi_+\}}{1-|t_{L}|^2|r_{n}|^2\exp\{i \chi_+\}}C_{in}^{(l,n)}\nonumber \\
C_{out}^{(r, n)}=r_L^{\star}\frac{t_n^{\star}\exp\{i
\chi_+\}}{1-|t_{L}|^2|r_{n}|^2\exp\{i \chi_+\}}
C_{in}^{(l,n-1)}\nonumber \\
+t_L^{\star}\frac{t_n^{\star}\exp\{i
\chi_+\}}{1-|t_{L}|^2|r_{n}|^2\exp\{i \chi_+\}}C_{in}^{(l,n)}
\label{AB}
\end{eqnarray}
%
%
where $n=1$ and
$$|t_L|^2 =\frac{\ell_c \omega_c \pi |\tilde{\Delta}|^2}{\alpha
v_1 \sqrt{\varepsilon-\varepsilon_0^{(0)}}} = \frac{\pi
|\Delta|^2}{\hbar \omega^\star V_1 P_0^\star} $$
 is the standard
Landau-Zener probabilities of the interband transitions that take
place at $p_y=0$ where two semiclassical trajectories approach
each other, $V_1 =d E_1/dp_y$ ,$\omega_H^\star =e H \hbar/m^\star
c$, $m^\star =m/\alpha^2$ , while $P_0^\star =\sqrt{2 m^\star
(E-E_0^{(0)})}$ and $\chi_+$ is the phase gain along the large
close orbit in Fig.\ref{trajectories}b (which is closed via the
small closed orbits, the latter being neglected), $|r_n|
=1-\frac{|t_1(\eta_n |^2}{2}|$
and
%
%
\begin{eqnarray}
 t_{1}
=-i\exp\{i \Theta\}\overline{\gamma}_1 \overline{w}_1 a_1
 \label{ProbabilityN1}
\end{eqnarray}
%
Here $\gamma_1, \;\overline{w}_1, \; a_1,\;\eta_1$ are defined in
Eqs.(\ref{Parameters},\ref{fw},\ref{a1}) in which $v_{1,2}^{(1)}
\equiv v_{0,1} = d \overline{\varepsilon}_{0,1}(p_y)/dp_y$ are the
velocities of the edge states $n=0$ and $n=1$.

As in the case under consideration  $E -E_0^{(0)}\sim \hbar
\omega_H$  and $V_1 \sim \sqrt{\hbar \omega_H/m}$ one has
$$|t_L|^2/|t_n|^2\sim (\frac{\alpha v_F}{\ell_c \omega_c})^{1/3}/ \sqrt{\varepsilon
-\varepsilon_0^{(0)}} \sim \alpha^{1/3} \ll 1$$ and hence  one may
approximately write
%
%
\begin{eqnarray}
  C_{out}^{(r,n-1)}\approx -t_n^\star \exp\{i \chi_+\}C_{in}^{(n)} \nonumber \\
 C_{out}^{r,n)}\approx +t_n \exp\{i \chi_+\}C_{in}^{(n-1)};
 \label{ABapprox}
\end{eqnarray}
%
%
From here it follows that in the first approximation in
$\overline{\gamma}_n \ll 1$ the probability for an electron
to pass from $p_y \rightarrow -\infty$ to $p_y \rightarrow
+\infty$ is approximately equal to $|t_n|$ for each incoming open
mode $0$ and $1$. Considering the electron energy at higher values
and checking possible semiclassical paths  of transmission one
finds that an analogous rule is correct for any number of incoming
open modes which may be written as follows:
%
%
\begin{eqnarray}
  |C_{out}^{(r,k)}|^2= D_{k}^{n -k}|C_{in}^{(l,n -k)}|^2; \;\;k=0,1,...n
 \label{Tn}
\end{eqnarray}
%
%
where $C_{in}^{(l,n}$ and $C_{out}^{(r,n)}$ are the amplitudes of
 the incoming (at $p_y
\rightarrow -\infty$) and the outgoing (at $p_y \rightarrow
+\infty$)  semiclassical wave functions
Eq.(\ref{FuctionsPInfinity}), respectively, $n$ is the maximal
quantum number of the incoming (outgoing) modes present at a given
energy $\varepsilon$ (e.g., $n=2$ for the case that $\varepsilon
\approx \varepsilon_0^{(2)}$, see Fig\ref{trajectories}). The
transmission probability is
%
%
\begin{eqnarray}
D_{k}^{n -k} =|\overline{\gamma}_k^{(n -k)}
\overline{w}_k^{(n -k)} a_1(\eta_n^{(n -k)})|^2;\nonumber \\
\;\;k=0,1,...n
 \label{TnDetail}
\end{eqnarray}
%
%
where $\overline{\gamma}_k^{(n -k)}\equiv
\overline{\gamma}_n,\;\;\overline{w}_k^{(n -k)}\equiv
\overline{w}_n , \;\;\eta_n^{(n_m -n)} \equiv \eta_n $ are
presented in Eqs.(\ref{Parameters},\ref{fw}) in which the Landau
numbers of the velocities  are now explicitly written:
%
%
\begin{eqnarray}
\overline{\gamma}_k^{(n -k)}=[\frac{\ell_c \omega_c}{\alpha
\overline{v}_n^{(n -k)}}]^{2/3}\Delta_{n }; \;\;
\overline{v}_k^{(n_m -k)} =\frac{v_k v_{n -k}}{v_k +v_{n -k}}; \nonumber \\
\eta_k^{(n -k)}=[\frac{\ell_c \omega_c}{\alpha \overline{v}_k^{(n
-k)}}]^{2/3}(\varepsilon_0^{(n )}-\varepsilon); \;\;
\overline{w}_k^{(n -k)}=\frac{\sqrt{v_k v_{n -k}}}{v_k+ v_{n -k}}.
 \label{ParameterDetail}
\end{eqnarray}
%
%

\emph{\textbf{Using Eq.(\ref{TnDetail}) and
Eq.(\ref{ParameterDetail}) at $n=k=0$ one gets Eqs.(10,11) of the
main text.}}

\section{Derivation of Eq.(12) of the main text.}

\emph{1 step, subsection A. Using the semiclassical wave functions
in the momentum space  (found in Section II) we find the electron
wave functions in the coordinate space which are the solution of
the Shr\;odinger Eq.(\ref{Shcroedinger}) written in theLandau
gauge ${\bf A}=(0,Hx,0)$ in which the project momentum $p_x$ does
not conserve and hence these functions can not be directly used to
describe the electron transport in terms of the incoming and
outgoing electrons.}

\emph{2 step. In subsection B we re-write the electron wave
functions in the form that is the solution of the Shr\" odinger
equation  in which the gauge ${\bf A}=(-Hy,0,0)$ is used that
allows to use the Landauer-B\"{u}ttiker approach and find Eq.(12)
of the main text. }

\subsection{Asymptotic of the  electron  wave functions at $x \gg
R_L $ and the gauge ${\bf A}=(0,Hx,0)$.}

 Equation (\ref{Tn}
)allows to connect  the asymptotic of the electron wave functions
both in
 the momentum and coordinate spaces for any number of incoming modes
 $n$.

 Taking  the quasiclassical functions Eqs.(\ref{SupplSemiclassic1},\ref{SupplSemiclassic2})
 in the region  $|p_y| \gg (e H/c) R_L = p_E =\sqrt{2m E}$ (that is far from the barrier
  where the electron spectrum is discrete,  $E_n = \hbar \omega (n +1/2)$),   one finds
  the Fourier factor  $Q $ (see Eq.(\ref{Q})   as follows:
%
%
\begin{eqnarray}
Q^{(l,r)}(x,p_y)=\sum_{\{n\}} \frac{1}{\sqrt{p_x^{(n)}}}\times \hspace{3.5cm} \nonumber \\
 \Big[C_{in}^{(l,r;n)}
\exp{\left(i\frac{p_x^{(n)}p_y}{p_1^2}\right)}
+ C_{out}^{(l,r;n)}\exp{\left(-i\frac{p_x^{(n)}p_y}{p_1^2}\right)}\Big] \nonumber \\
\times \exp{\left[-\frac{(x+c p_y/eh)^2}{2
\ell_c^2}\right]}H_n(\frac{(x+c p_y/eh)}{\ell_c}) \label{Q2}
\end{eqnarray}
%
%
where the sum is over all edge state  modes $n$ present at a given
electron energy $\varepsilon$;  $H_n(x)$ is the Hermite
polynomial; the superscripts $(l)$ and $(r)$ denote the regions
$p_y < 0$ and $p_y
> 0$, respectively. While writing Eq.(\ref{Q2}) we have taken into
account Eq.(\ref{Suppledge0}) and the fact that $ R_n^{(r)}
\approx 0$ at $p_y < 0$, and $R_n^{(l)} \approx 0$ at $p_y > 0$.

Inserting Eq.(\ref{Q2}) in Eq.(2)of the main text one finds the
asymptotic of the wave function $\Psi(x,y)^{(l,r)}$ at $|x| \gg
R_L$ in the left part (l), $x < 0 $,  and the right part (r), $x <
0 $, of the system as follows:
%
%
\begin{eqnarray}
\Psi^{(l)}(x,y)=  \hspace{6cm} \nonumber \\
\exp\{-i\frac{eH\hbar}{c}xy\} \sum_{\{n\}} \Big[C_{in}^{(l,n)}
\exp{\left(-i\frac{e H}{c p_1^2}p_x^{(n)} x\right)}\times \nonumber \\
 \int_{-\infty}^{+\infty}  \exp\{-q^2/2\sigma\}
 \exp\{i\left(\frac{p_x^{(n)}}{p_1^2} + \frac{y}{\hbar}\right)q\}
 H_n(\frac{q}{\sqrt{\sigma}}) dq     \nonumber \\
+C_{out}^{(l,n)}\exp{\left(-i\frac{e H}{c p_1^2}p_x^{(n)}
x\right)}
 \int_{-\infty}^{+\infty} \exp\{-q^2/2\sigma\} \nonumber \\
  \times
  \exp\{-i\left(\frac{p_x^{(n)}}{p_1^2}- \frac{y}{\hbar}\right)q\}
 H_n(\frac{q}{\sqrt{\sigma}}) dq\Big]
\label{Psi1a}\hspace{0.2cm}
\end{eqnarray}
%
and
%
%
\begin{eqnarray}
\Psi^{(r)}(x,y)=  \hspace{6cm} \nonumber \\
\exp\{-i\frac{eH\hbar}{c}xy\} \sum_{\{n\}} \Big[C_{out}^{(r,n)}
\exp{\left(-i\frac{e H}{c p_1^2}p_x^{(n)} x\right)}\times \nonumber \\
 \int_{-\infty}^{+\infty}  \exp\{-q^2/2\sigma\}
 \exp\{i\left(\frac{p_x^{(n)}}{p_1^2} + \frac{y}{\hbar}\right)q\}
 H_n(\frac{q}{\sqrt{\sigma}}) dq     \nonumber \\
+C_{in}^{(r,n)}\exp{\left(-i\frac{e H}{c p_1^2}p_x^{(n)} x\right)}
 \int_{-\infty}^{+\infty} \exp\{-q^2/2\sigma\} \nonumber \\
  \times
  \exp\{-i\left(\frac{p_x^{(n)}}{p_1^2}- \frac{y}{\hbar}\right)q\}
 H_n(\frac{q}{\sqrt{\sigma}}) dq\Big]
\label{Psi1b}\hspace{0.2cm}
\end{eqnarray}
%
where $p_x^{(n)}=\sqrt{2 m [E-\hbar \omega_H(n+1/2)]}, \;\;
p_1^2=m \hbar \omega_1, \;\; \sigma=e \hbar H/c$.

In the next section we show that these wave functions may be
presented in the form convenient for calculations of the current
flux through the point contact.

\subsection{Gauge change in the  presentation of the found wave functions. }

\emph{The found wave functions Eq.(\ref{Psi1a},\ref{Psi1b}), are
the solutions of the Schr\"odinger equation Eq.(1) of the main
text in which the gauge ${\bf A}=(0,Hx,0)$ is used.   In this
gauge the
 $p_x$-projection of the momentum of the incoming and outgoing
 edge states does not conserve and hence one can not directly use
 these functions to find the current flowing along the
point contact. In this subsection we recast the wave functions
Eq.(\ref{Psi1a},\ref{Ps2a}) into the form that corresponds to the
gauge ${\bf A}=(-Hy,0,0)$ in which $p_x$ conserves.}

\subsubsection{Derivation of the equality that allows the gauge transformation.}
We use the generating function of the Hermite polynomials (see,
e.g., \cite{Gradshtein})
\begin{eqnarray}
\exp\{-s^2+2 s q\}=\sum_{k=0}^{\infty}\frac{s^k}{k !}H_k(q);
 \label{generating function}
\end{eqnarray}
Multiplying the both sides of it by $\exp\{-q^2+i\widetilde{y}q\}$
and integrating with respect to q one gets
\begin{eqnarray}
\int_\infty^\infty\exp\{-q^2+i\widetilde{y}q\} \exp\{-s^2+2 s q\}dq\nonumber \\
=\sum_{k=0}^{\infty}\frac{t^k}{k
!}\int_\infty^\infty\exp\{-q^2+i\widetilde{y}q\}  H_k(q)dq;
 \label{generating function1}
\end{eqnarray}
Integrating the left-hand side one easily finds
\begin{eqnarray}
\sum_{k=0}^{\infty}\frac{s^k}{k !}\int_\infty^\infty\exp\{-q^2+i\widetilde{y}q\}  H_k(q)dq\nonumber \\
=\sqrt{2 \pi}e^{-\widetilde{y}^2/2}\exp\{-(is)^2+2
(is)\widetilde{y}\};
 \label{generating function2}
\end{eqnarray}
Applying Eq.(\ref{generating function}) to the right-hand side of
this equation one  gets the following:
\begin{eqnarray}
\sum_{k=0}^{\infty}\frac{s^k}{k !}\Big\{\int_\infty^\infty  \exp\{-q^2+i\widetilde{y}q\}  H_k(q)dq\nonumber \\
-i^k \sqrt{2 \pi}H_k(\widetilde{y})\Big\}=0 ;
 \label{generating function2}
\end{eqnarray}
As $s$ is arbitrary one gets the following equation:
\begin{eqnarray}
\int_\infty^\infty  \exp\{-q^2+i\widetilde{y}q\}  H_k(q)dq =i^k
\sqrt{2 \pi}H_k(\widetilde{y})\}
 \label{generating function3}
\end{eqnarray}

\subsubsection{Wave functions of the incoming electrons  and  those scattered by the lateral junction
in which $p_x$ conserves.}

Now applying the obtained equality Eq.(\ref{generating function3})
to Eqs.(\ref{Psi1a},\ref{Psi1b}) we get the wave functions as
follows:
%
\begin{eqnarray}
x \rightarrow -\infty; \hspace{5cm}\nonumber \\
\Psi^{(l)}(x,y)=\sqrt{2\pi \sigma} e^{-i (eH/\hbar c)xy}\times \hspace{2.5cm}  \nonumber \\
 \sum_{\{n\}} i^n \Big[ C_{in}^{(l,n)}\exp\{-i\frac{p_x^{(n,
\star)} x}{\hbar}\} \exp\{-\frac{y_{n,+}^2}{2 \ell_c^2} \}
 H_n(\frac{y_{n,+}}{\ell_c})     \nonumber \\
+ C_{out}^{(l,n)} \exp\{+i\frac{p_x^{(n,\star)} x}{\hbar}\}
\exp\{-\frac{y_{n,-}^2}{2 \ell_c^2} \}
 H_n(\frac{y_{n,-}}{\ell_c})  \Big]
 \label{Psi2}
\end{eqnarray}
%
%
and
%
\begin{eqnarray}
x \rightarrow +\infty; \hspace{5cm}\nonumber \\
\Psi^{(r)}(x,y)=\sqrt{2\pi \sigma} e^{-i (eH/\hbar c)xy}\times  \hspace{2.5cm}\nonumber \\
 \sum_{\{n\}} i^n \Big[ C_{out}^{(r,n)}\exp\{-i\frac{p_x^{(n,
\star)} x}{\hbar}\} \exp\{-\frac{y_{n,+}^2}{2 \ell_c^2} \}
 H_n(\frac{y_{n,+}}{\ell_c})     \nonumber \\
+ C_{in}^{(r,n)} \exp\{+i\frac{p_x^{(n,\star)} x}{\hbar}\}
\exp\{-\frac{y_{n,-}^2}{2 \ell_c^2} \}
 H_n(\frac{y_{n,-}}{\ell_c})  \Big]
 \label{Psi3}
\end{eqnarray}
%
%
where $p_x^{(n, \star)}= \sqrt{2 m^\star [E - \hbar
\omega_H(n+1/2)]}$, $m^\star =m/ \alpha^2$ is the re-normalized
mass and $y_{n,\pm} = y \pm c p_x^{n,\star}/e H$ while the
constant factors at the incoming, $C_{in}^{(r,n)}$, and outgoing
wave functions, $C_{out}^{(r,n)}$, (which are normalized to the
unity flux) are connected to one another by the relations
Eqs.(\ref{Tn},\ref{TnDetail},\ref{ParameterDetail}).

The wave functions inside the square brackets are solutions of
Schr\"odinger's equation Eq.(\ref{Shcroedinger}) at $|x| \gg R_L$
in which the gauge of the vector potential is changed to  ${\bf A}
= (-Hx, 0, 0)$ that conserves the projection of the electron
momentum $p_x$ parallel to the longitudinal direction of the
junction.

Using Eqs.(\ref{Psi2},\ref{Psi3})together with   Eq.(\ref{Tn}) one
finds the current flowing along the junction as follows:
\begin{eqnarray}\label{current}
I(V) = \frac{2 e}{h} \sum_{n=0}^{\infty}\sum_{k=0}^{n}
\int_{-\infty}^{\infty} d E D_n^{n-k}\big(E-E_{n}^{n-k}(V)\big)
\\\nonumber [f(E+eV/2)- f(E-e V/2)]~,
\end{eqnarray}
where $E_{n}^{n-k}(V)$ is the energy of degeneration of the left
and right edge states under applied voltage drop $V$, that is $
E_{n}^{n-k}(V)\equiv E_n^{(l)}(p_y)+eV/2 =E_k^{(r)}(p_y)-eV/2$.

\emph{\textbf{Eq.(12) of the main text is a simplified form of
Eq.(\ref{current}).}}

\end{document}